\documentclass[aps,prl,twocolumn,showpacs,groupedaddress]{revtex4}
\usepackage{graphicx}
\usepackage{amsfonts}
\usepackage{amsmath}
\usepackage{amssymb}
\usepackage{bm}

\begin{document}

\title{Floquet Weyl Semimetal Induced by Off--Resonant Light}
\author{Rui Wang$^{1}$, Baigeng Wang$^{1}$, Rui Shen$^{1}$, L. Sheng$^{1}$,
D. Y. Xing$^{1}$ and Sergey Y. Savrasov$^{1,2}$}
\affiliation{$^1$National Laboratory of Solid State Microstructures and School of
Physics, Nanjing University, Nanjing 210093, China\\
$^2$Department of Physics, University of California, Davis, California
95616, USA}
\date{\today }

\begin{abstract}
We propose that a Floquet Weyl semimetal state can be induced in
three--dimensional topological insulators, either nonmagnetic or magnetic,
by the application of off--resonant light. The virtual photon processes play
a critical role in renormalizing the Dirac mass and so resulting in a
topological semimetal with vanishing gap at Weyl points. The present
mechanism via off--resonant light is quite different from that via
on--resonant light, the latter being recently suggested to give rise to a
Floquet topological state in ordinary band insulators.
\end{abstract}

\maketitle

\emph{Introduction}--Topological states of matter, such as two--dimensional
(2D) quantum spin Hall insulators or 3D topological insulators (TIs)~\cite%
{Kanea,Kaneb,ShouChenga,ShouChengb,Fua,ZhangH,Balentsa,Fub}, have received a
great interest in recent years. Searching for such states in solid state
materials~\cite{ZhangH,Bernevig,Hsieh} has partly gained success; however,
candidate materials for TIs are still very limited. Inspiringly, an
intriguing method was put forward to realize topologically non--trivial
phases in non--equilibrium by applying time--dependent perturbations to
trivial phases~\cite{Kitagawa,Gu,Takuya,Netanel,Eric,Niu}. 
Typical examples are the optically--activated anomalous Hall effect and spin
Hall effect in \emph{n}--doped paramagnetic semiconductors~\cite{Niu}, and
the so--called Floquet topological insulator (FTI) suggested by Lindner,
Refael, and Galitski~\cite{Netanel}, whose quasi--energy spectrum exhibits a
single pair of helical edge states due to the  on--resonant--light-- induced
band inversion. In the bulk FTI spectrum there is an avoided crossing
separating the reshuffled valence band from the conduction band.

Recently, another type of topological phase termed the Weyl semimetal (WSM)~%
\cite{Wan} has attracted much attention. The WSM has no bulk gap but enjoys
gapless nodes distributed in momentum space, and it possesses chiral surface
states and Fermi arcs terminating at Weyl points with opposite chirality.
Due to its remarkable electromagnetic properties such as the anomalous Hall
effect~\cite{Yang,Balentsb} and chiral magnetic effect~\cite{Kenji}, the WSM
is regarded as a promising candidate for future applications in spintronics.
Unfortunately, despite several proposals have been put forward in various
systems~\cite{Wan,Balentsb,Cho}, the WSM has not been realized
experimentally yet. As a result, theoretical proposals on the WSM state that
are feasible for experimental realization are still highly desirable.


\begin{figure}[tbp]
\includegraphics[width=3.5in]{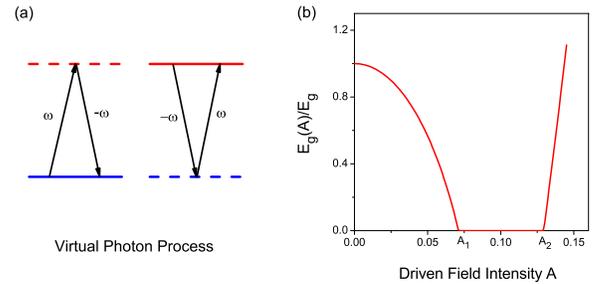}
\caption{(color online) (a)Schematic diagram for virtual photon processes.
(b)Renormalized energy gap as a function of light field intensity. Here $%
A_{1}$ and $A_{2}$ are the thresholds (in unit of $\text{{\AA }}^{-1}$), at
which a pair of Weyl points are generated and annihilated, respectively.}
\end{figure}
In this Letter, we show that a non--equilibrium WSM state can be induced in
3D topological insulators by the application of off--resonant light, using
the Floquet picture. In contrast to the on--resonant optical induction in
Ref.~\cite{Netanel}, we focus on the off--resonant effect of light on band
structures, and the underlying physics is outlined as follows. Considers a
two--band system with bulk energy gap $E_{g}$ under the application of
off--resonant light. For light with off--resonant frequency $\omega $, the
real process of photon absorption/emission cannot occur because of
limitation of energy conservation condition. However, the off--resonant
light can affect the electron system via virtual photon processes, e.g., two
second--order processes shown in Fig.~1(a), where electrons absorb and then
emit a photon and electrons first emit and absorb a photon. Such virtual
photon processes have at least two effects. First, they can break the time
reversal symmetry (TRS) if the driven field is nonlinearly polarized, which
is one of essential conditions to realize WSM. Second, they can generate a
self--energy $\mathbf{\Sigma }$ that renormalizes the Dirac mass and so bulk
energy gap $E_{g}(A)$ of the electron system. It will be shown that $E_{g}(A)
$ first decreases with light field amplitude $A$ and vanishes at threshold $%
A_{1}$, as shown in Fig.~1(b). Once the light field amplitude is increased
to another threshold $A_{2}$, the bulk gap reopens and $E_{g}(A)$ increases
with further increasing $A$. For $A_{1}<A<A_{2}$, the bulk energy spectrum
is gapless at Weyl points, and the system will be shown to be a Floquet WSM.

\emph{Floquet theory on topological insulator}--We start from a modified
Dirac Hamiltonian~\cite{Shun}, which was used to describe magnetic 3D TI, 
\begin{equation}
H(\mathbf{k})=\sum_{i=1}^{3}ak_{i}\Gamma _{i}+M\Gamma _{4}+m\Gamma ^{\prime
}.  \label{eq2}
\end{equation}%
The first two terms describe the nonmagnetic 3D TI, where $\Gamma _{i}$ ($%
i=1,...,4$) are the Dirac matrices that satisfy Clifford algebra in
Euclidean space: $\{\Gamma _{i},\Gamma _{j}\}=\delta _{ij}$. The Dirac mass
is given by $M=M_{0}+bk^{2}$. It is well known that $M_{0}b<0$ guarantees a
topologically nontrivial phase while $M_{0}b>0$ suggests a normal insulator
(NI). Once $\Gamma _{1},...,\Gamma _{4}$ are specified, the other Dirac
matrices can be constructed as $\Gamma _{0}=\mathbf{1}_{4\times 4}$, $\Gamma
_{5}=\prod_{i=0}^{4}\Gamma _{i}$, and $\Gamma _{ij}=-\frac{i}{2}[\Gamma
_{i},\Gamma _{j}]$. The last term represents a time-reversal-breaking
exchange field with $\Gamma ^{\prime }$ satisfying $[\Gamma ^{\prime
},\Gamma _{4}]=0$. A competition between $m\Gamma ^{\prime }$ and $M\Gamma
_{4}$ will enable the surface Dirac fermions to acquire mass, leading to a
quantum anomalous Hall (QAH) effect. Since the choice of $\Gamma^\prime$ is
flexible, we choose $\Gamma ^{\prime }=\Gamma_{12}$ for the purpose of
clarity.

An in--plane polarized electric field can be expressed by its vector
potential $\mathbf{A}=(A_{x}\sin \omega t,A_{y}\sin (\omega t+\phi ),0)$
with $\omega $ as the frequency of the electric field and $\phi $ describing
the polarization direction. For a spatially uniform electromagnetic field,
the effect on the spin degree of freedom may be negligible while the effect
on the orbital degree of freedom can be included through the substitution: $%
H(\mathbf{k})\rightarrow H(\mathbf{k}+\mathbf{A}(t))$, yielding 
\begin{equation}
H(\mathbf{k},t)=H(\mathbf{k})+\mathbf{V(t)}\cdot \mathbf{\Gamma }.
\label{eq10}
\end{equation}%
Here the second term is the time--dependent perturbation induced by the
driven field with $\mathbf{\Gamma }=(\Gamma _{1},$ $\Gamma _{2},\Gamma
_{3},\Gamma _{4})$ and $\mathbf{V(t)}=(aA_{x}\sin \omega t,aA_{y}\sin
(\omega t+\phi ),$ $0,$ $bA_{x}^{2}\sin ^{2}\omega t+bA_{y}^{2}\sin
^{2}(\omega t+\phi )+2bk_{x}A_{x}\sin \omega t+2bk_{y}A_{y}\sin (\omega
t+\phi ))$.

Due to the discrete time--translational invariance of $H(\mathbf{k},t)$, the
eigenvectors can be written as $|\psi _{\alpha }(t)\rangle =e^{-i\epsilon
_{\alpha }t}|\phi _{\alpha }(t)\rangle $, where $|\phi _{\alpha }(t)\rangle
=|\phi _{\alpha }(t+T)\rangle $ is the Floquet state with $T=2\pi /\omega $
and $\epsilon _{\alpha }$ is the quasi--energy restricted in the range of $%
-\omega /2<\epsilon _{\alpha }<\omega /2$. In the framework of the Floquet
theory, the time--dependent Hamiltonian Eq.~\eqref{eq10} can be mapped onto
a time--independent Floquet operator $\mathbf{F}=\mathbf{\Omega }\otimes 
\mathbf{H}_{4\times 4}$ in the Sambe space~\cite{Sambe}, with $\mathbf{%
\Omega }$ as an infinite--dimensional matrix in the Floquet band space and $%
\mathbf{H}_{4\times 4}$ in the basis of $H(\mathbf{k})$ of the undriven
system. The matrix element of the Floquet operator is given by 
\begin{equation}
\mathbf{F}_{m,n}(\mathbf{k})=H(\mathbf{k})+\mathbf{\Gamma }\cdot \mathbf{V}%
_{m,n}-n\omega \delta _{nm},  \label{eq3}
\end{equation}%
where $\mathbf{V}_{m,n}=\frac{1}{T}\int_{-T/2}^{T/2}dt\mathbf{V(t)}%
e^{-i(n-m)\omega t}$, and $m$ and $n$ denote the indices of Floquet bands.
In principle, an exact diagonalization of the Floquet operator can give all
the quasi--energy spectra. In order to obtain analytic results, we perform a
perturbation theory, which holds for $a^{2}A_{x}A_{y}/M\omega \ll 1$~\cite%
{Ruif}, to obtain the effective Floquet operator~\cite{Kitagawa}: $%
F_{eff}=F_{0,0}+[F_{-1,0},F_{1,0}]/\omega $. Both terms $F_{0,0}$ and $%
[F_{-1,0},F_{1,0}]/\omega $ arise from the second--order virtual photon
processes shown in Fig.\ 1(a), where a photon is first absorbed (released)
and then released (absorbed). As will be demonstrated below, $%
[F_{-1,0},F_{1,0}]/\omega $ breaks the TRS for a nonlinearly--polarized
electric field, while $F_{0,0}$ differs from the undriven Hamiltonian $H(%
\mathbf{k})$ by a modification $\mathbf{\Sigma }$ that renormalizes the
Dirac mass and generates Weyl points.

Using the perturbation theory, we obtain the effective Floquet operator
around the $\Gamma $ point in momentum space as 
\begin{equation}
F_{eff}(\mathbf{k})=\sum_{i=1}^{3}ak_{i}\Gamma _{i}+\widetilde{M}\Gamma _{4}+%
\widetilde{m}\Gamma ^{\prime },  \label{eq12}
\end{equation}%
for the driven system, where $\widetilde{m}=m+a^{2}(A_{x}A_{y}/\omega )\sin {%
\phi }$ and $\widetilde{M}=\widetilde{M}_{0}+bk^{2}$ with $\widetilde{M}%
_{0}=M_{0}+b(A_{x}^{2}+A_{y}^{2})/2$. Comparing Eq.~\eqref{eq12} with Eq.~%
\eqref{eq2} for the undriven system, one finds two important differences
between them. First, the renormalized exchange field is enhanced due to term 
$[F_{-1,0},F_{1,0}]/\omega $. For a nonlinearly--polarized field ($\phi \neq
0$), the TRS is broken even though there is no exchange field ($m=0$).
Second, the Dirac mass is dressed due to term $F_{0,0}$. Since the TI is
chosen to be the initial state with $M_{0}>0$ and $b<0$, $\widetilde{M}_{0}$
is effectively reduced. Since $\widetilde{m}>m$ and $\widetilde{M}<M$, it is
expected that the competition between $\widetilde{m}\Gamma ^{\prime }$ and $%
\widetilde{M}\Gamma _{4}$ is favorable to emergence of the QAH phase and
even WSM phase. If the field intensity is large enough to make $\widetilde{M}%
_{0}$ change its sign from positive to negative so that $\widetilde{M}_{0}b>0
$, the system can be driven to be a topologically trivial insulator phase.

To be more explicit, we now represent the Dirac matrices as $(\Gamma
_{1},\Gamma _{2},\Gamma _{3},\Gamma _{4})=(\tau ^{x}s^{x},\tau
^{x}s^{y},\tau ^{x}s^{z},-\tau ^{z}s^{0})$, where $\mathbf{s }$ ($\mathbf{%
\tau }$) is the Pauli matrix in spin (orbital) space. In this
representation, Eq.~\eqref{eq12} reads 
\begin{equation}
F_{eff}(\mathbf{k})=a\tau ^{x}(s ^{x}k_{x}+s^{y}k_{y}+s^{z}k_{z})-\widetilde{%
M}\tau ^{z}s ^{0}+\widetilde{m}\tau ^{0}s^{z}.  \label{eq5}
\end{equation}%
In the absence of the last term, Eq.~\eqref{eq5} is an isotropic version of
the model in Ref.~\cite{ZhangH}, which describes the 3D TI $\text{Bi}_{2}%
\text{Se}_{3}$. To see the possibility of the topological phase transition,
we first make a rotation in orbital space: $\tau ^{x}\rightarrow \tau ^{z}$
and $\tau ^{z}\rightarrow -\tau ^{x}$, and then a canonical transformation: $%
\tau ^{\pm }\rightarrow s^{z}\tau ^{\pm }$ and $s^{\pm }\rightarrow s^{\pm
}\tau ^{z}$. After $F_{eff}(\mathbf{k})$ in Eq.\ (5) is diagonalized in
orbital space, we have 
\begin{equation}
F_{\pm }(\mathbf{k})=a(s^{x}k_{x}+s^{y}k_{y})+s^{z}\Delta _{\pm }(\mathbf{k}%
),  \label{eq6}
\end{equation}%
with $\Delta _{\pm }(\mathbf{k})=\widetilde{m}\pm \sqrt{\widetilde{M}%
^{2}+a^{2}k_{z}^{2} }$. For a fixed $k_{z}$, Eq.~\eqref{eq6} is a 2D Dirac
Hamiltonian. It is well known that a sign change of the 2D Dirac mass
signals a topological phase transition, characterized by a change of the
first Chern number, which provides a method to distinguish different phases.
From $\Delta _{-}(0,0,k_{z})=0$, we have solution $k_z = \pm k_{zc}$ with $%
ak_{zc} =\sqrt{\widetilde{m}^{2}-\widetilde{M}^{2}}$. For $\widetilde{m}^2>%
\widetilde{M}^{2}$, there are two real roots $\pm k_{zc}$ for $k_{z}$, and $%
(0,0,\pm k_{zc})$ are just the two Weyl points in momentum space of the WSM.
Further more, an expansion of Floquet operator $F_{-}$ in Eq.~\eqref{eq6}
near the two gapless nodes $\pm k_{zc}$ yields 
\begin{equation}
F_{-}(\mathbf{k})=a\tau ^{x}q_{x}-a\tau ^{y}q_{y}\mp a\sqrt{\widetilde{m}%
^{2}-M_{0}^{2}}\tau ^{z}q_{z},  \label{eq7}
\end{equation}%
which is the same as the Hamiltonian defining the Weyl fermions with
chirality $C=\pm 1$. Here $bk^{2}$ in the Dirac mass is temporarily
neglected for clarity. On the other hand, for $\widetilde{m}^2<\widetilde{M}%
^{2}$, there exists no real root of $k_{zc}$ for $\Delta _{-}(0,0,k_{z})=0$
and so the quasi-energy spectrum is gapped throughout the momentum space. In
this case, either QAH or NI phase is possible, depending on whether $%
\widetilde{M}_{0}b<0$ or $\widetilde{M}_{0}b>0$. As a result, there is a
topological phase transition at $\widetilde{m}^2 = \widetilde{M}^{2}$ from
the WSM phase to either QAH or NI phase.

\begin{figure}[tbp]
\includegraphics[width=3.4in]{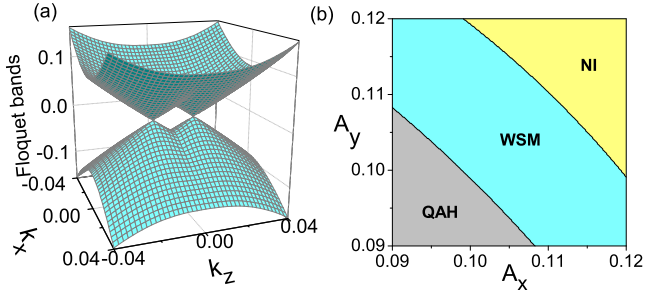}
\caption{(color online) (a) Calculated Floquet bands for a 3D TI subjected
to a circularly polarized light with $A_{x}=A_{y}=0.105\text{{\AA }}^{-1}$,
and (b) the corresponding topological phase diagram. The parameters are
chosen as $M_{0}=0.28eV$, $a=2.8eV\text{{\AA }}$, $b=-25.7eV\text{{\AA }}^{2}
$, and $\protect\omega =3eV$. }
\end{figure}

\begin{figure}[tbp]
\includegraphics[width=3.4in]{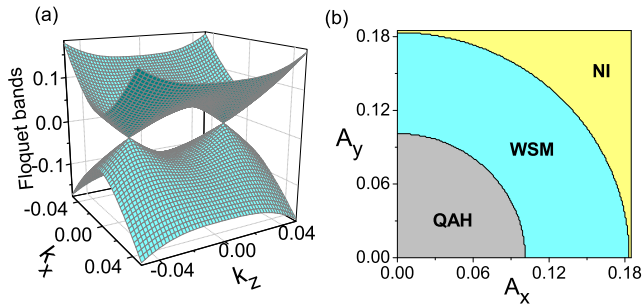}
\caption{(color online) (a) Calculated Floquet bands for a 3D magnetic TI
subjected to a linearly polarized light with $A_{x}=A_{y}=0.1\text{{\AA }}%
^{-1}$, and (b) the corresponding topological phase diagram. The exchange
energy is chosen as $m=0.15$ $eV$ and the other parameters are the same as
in Fig.~2(a). }
\end{figure}

From Eq.~\eqref{eq5} we have numerically calculated the bulk quasi--energy
dispersion in a single Floquet band for different polarizations. We first
consider the case of $m=0$ and $\phi =\pi /2$ (circularly polarized light).
It is found that with increasing the off--resonant light intensity, the
energy gap $E_{g}$ of the undriven TI becomes smaller and smaller, and
vanishes at $k_{zc}=0$. With further increasing $A$, a pair of Weyl points
at $\pm k_{zc}$ are separated from each other, as shown in Fig.~2(a). A
continued increase of $A$ can make two Weyl points merge into a single Dirac
point at $k_{zc}=0$, and then the gap reopens, resulting in a topologically
trivial insulator. Figure 2(b) is the calculated phase diagram, where the
WSM phase is in between the QAH and NI phases. Second, we consider the
linearly-polarized case of $\phi =0$ with $m\neq 0$ taken. Similar
topological phase transitions are obtained, as shown in Fig.~3. If $\phi =0$
and $m=0$, the system still has the TRS. In this case, there is no WSM phase
in the phase diagram, and the TI phase is transited to the NI phase under
the application of off--resonant light.

\emph{Lattice model}--In what follows we study a more realistic diamond
lattice model to support the above results. Its tight--binding Hamiltonian
is given by 
\begin{equation}
H=t\sum_{\langle ij\rangle }c_{i}^{\dagger }c_{j}+i(8\lambda
_{SO}/a^{2})\sum_{\langle \langle ij\rangle \rangle }c_{i}^{\dagger }\mathbf{%
s}\cdot (\mathbf{d}_{ij}^{1}\times \mathbf{d}_{ij}^{2})c_{j}.  \label{eq8}
\end{equation}%
This model was introduced to describe 3D TI~\cite{Fua}. Here we take $%
t_{1}=t+\delta t$ and $t_{2}=t_{3}=t_{4}=t$ with the distortion $\delta t$
along the $(111)$ direction. For $\delta t>0$, the system lies in the strong
topological insulator (STI) phase with $Z_{2}$ index as $\nu _{0}=1$, and $%
\delta t<0$ corresponds a weak topological insulator (WTI) phase with $\nu
_{0}=0$.

First, we do not consider Zeeman splitting and only focus on the linearly
polarized electric field $\mathbf{A}=(A_{x}\sin \omega t,A_{y}\sin \omega
t,0)$. Using the procedure same as above, we obtain the Floquet operator in
Sambe space as 
\begin{equation}
F_{m,n}(\mathbf{k})=\mathbf{\Gamma }\cdot \mathbf{D}_{q}(\mathbf{k})-n\omega
\delta _{mn},  \label{eq9}
\end{equation}%
where $m$ and $n$ denote different Floquet bands and $q=m-n$. The Dirac
matrices in this case are represented as $\mathbf{\Gamma }=(\tau
^{z}s^{x},\tau ^{z}s^{y},\tau ^{z}s^{z},\tau ^{x}s^{0},\tau ^{y}s^{0})$,
where $\tau $ and $s$ are the Pauli matrices associated with A-B sublattice
and true spin subspace, respectively; and $\mathbf{D}_{q}(\mathbf{k}%
)=(d_{1q}(\mathbf{k}),d_{2q}(\mathbf{k}),d_{3q}(\mathbf{k}),d_{4q}(\mathbf{k}%
),d_{5q}(\mathbf{k}))$, where $d_{iq}(\mathbf{k})$ are expressed via the $q$%
th order Bessel functions, which can be obtained by first Fourier
transforming Eq.~(\ref{eq8}) and then mapping it onto the Floquet operator
in Sambe space. In the off--resonant region of $\omega >8t$, we set $m=n=0$
in Eq.~\eqref{eq9} to obtain $F_{0,0}$, which is the most relevant
contribution from all the virtual photon processes.

In Eq.~\eqref{eq9} for $q=0$, both the TRS and the inversion symmetry, which
can be represented as $\hat{\Theta}=is^{y}K$ and $\hat{P}=\tau ^{x}s^{0},$
respectively~\cite{Fub}, remain unchanged. Therefore, the $Z_{2}$ index of
the driven Hamiltonian can be conveniently calculated, which is modulated by
the off--resonant light. Figure 4(a) shows the phase diagram, in which there
is a topological phase transition between the STI and WTI phases.

\begin{figure}[tbp]
\includegraphics[width=3.4in]{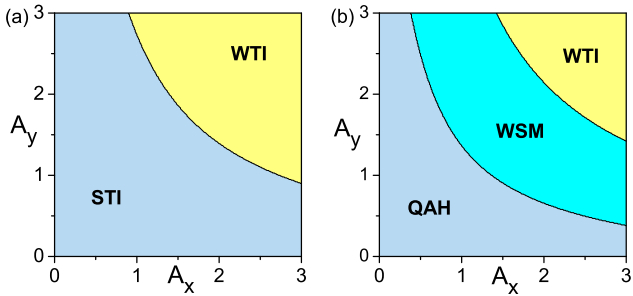}
\caption{(color online) Calculated topological phase diagrams of the diamond
lattice model driven by a linearly polarized light ($\protect\phi =0$) with $%
m=0$ (a) and $m=0.2eV$ (b). The parameters are taken as $t_{2,3,4}=1$, $%
t_{1}=1.4$, and $\protect\lambda _{SO}=0.05$. }
\label{fig4}
\end{figure}

Next, we take the Zeeman splitting into account with term $m\tau ^{0}s^{z}$
included in Eq.~\eqref{eq9}. The eigenvalues are obtained as $E_{\pm
}^{2}=d_{1}^{2}+d_{2}^{2}+(\sqrt{d_{3}^{2}+d_{4}^{2}+d_{5}^{2}}\pm m)^{2}.$
Note that there are gapless nodes only for $E_{-}$, which requires $%
d_{1}=d_{2}=0$ and $m=\sqrt{ d_{3}^{2}+d_{4}^{2}+d_{5}^{2}}$. There are
three equations that determine three variables $k_{x}$, $k_{y}$, and $k_{z}$%
. The solutions for them are exactly the Weyl nodes in momentum space. To
make this argument more convincing, we expand the Floquet operator around
point $X^{z}=(0,0,2\pi )$ and then make the canonical transformation $s^{\pm
}\rightarrow \tau ^{z}s^{\pm }$ and $\tau ^{\pm }\rightarrow \tau ^{\pm
}s^{z}$ to obtain the effective Hamiltonian as 
\begin{equation}
H_{eff}=\lambda _{SO}(s^{x}q_{x}-s^{y}q_{y})+s^{z}\Delta ^{\pm }(\mathbf{q};%
\mathbf{A}).  \label{eq11}
\end{equation}%
Here the effect of the light field $\mathbf{A}$ has been absorbed into the
Dirac mass: $\Delta ^{\pm }(\mathbf{q};\mathbf{A})=m\pm \lbrack
M_{z}^{2}+h_{ab}q_{a}q_{b}]^{1/2}$ with $M_{z}=\sum_{i=1}^{4}sgn(\mathbf{%
e_{i}}\cdot \mathbf{\hat{z}})$, where $\mathbf{e_{i}}$ is the nearest bond
vector of the diamond lattice and $h_{ab}$ is a tensor describing the
anisotropy that is determined by field parameters $A_{x}$ and $A_{y}$.
Equation \eqref{eq11} is of the same form as Eq.~\eqref{eq6}, so that a
similar analysis can be made here. The numerical results searching for
different phases are shown in Fig.\ 4(b), in which the WSM phase is in
between the QAH and magnetic WTI phases.

\emph{Experimental realization and conclusion}--First, in order to take
advantage of the off--resonant process, the frequency of the electric field
should be of the order of $10^{3}$ $\text{THz}$. Therefore, an ultraviolet
light is qualified. Second, the field intensity has to cross the topological
phase transition line as shown in Fig.~3, i.e., $A_{0}>0.1\text{{\AA }}^{-1}$%
. In terms of the electric field amplitude $E_{0}$, it should be of the
order of $0.1V\text{{\AA }}^{-1}$. It needs to be stressed that the
amplitude threshold will be greatly reduced for a TI material with a smaller 
$M_{0}$ and a larger $|b|$. For example, $E_{0}$ is reduced to $10^{-4}V%
\text{{\AA }}^{-1}$ for $M_{0}\sim 0.01eV$ and $b\sim -100eV\text{{\AA }}^{2}
$.

In conclusion, we have shown that the Floquet WSM phase can be induced in 3D
TIs by the use of the off--resonant light. The virtual photon
absorption/emission processes play a key role in renormalizing the Dirac
mass, in closing the bulk gap, as well as in breaking the TRS, resulting in
a nontrivial WSM phase. From both the continuous and lattice models of the
Floquet theory, very similar phase diagrams have been obtained. With
increasing the light intensity, the QAH phase first transits to the WSM
phase and then to the NI or magnetic WTI phase. According to recent
experiments such as the observation of the QAH effect in magnetic TIs~\cite%
{Chang} and the realization of Floquet topological insulators in a photonic
system~\cite{Mikael}, the present proposal of the Floquet WSM state can be
realized in experiments. In addition, we wish to point out that the
generalized Dirac Hamiltonian, Eq.~\eqref{eq2}, can describe not only the 3D
TIs, but also the other systems such as the polyacetylene, $p$--wave pairing
superconductor, and $^{3}\text{He}$-$\text{A}$ and $^{3}\text{He}$-$\text{B}$
phases~\cite{Shun}. Therefore, it is expected that the light field may also
induce topological phase transitions in those systems.

\begin{acknowledgments}
This work is supported by the State Key Program for Basic Research of China
under Grant No. 2011CB922103, and by the National Natural Science Foundation
of China under Grants No. 60825402, No. 11023002, and No. 91021003.
\end{acknowledgments}


\begin{thebibliography}{99}
\bibitem{Ruif} This condition is not nessesary in the lattice model, because
the vector potential enters into $F_{m,n}$ through ($m-n$)--th order Bessel
functions, which decay oscillatorily.
\end{thebibliography}
\end{document}